# Coexistence of multi-scale domains in ferroelectric polycrystals with non-uniform grain-size distributions

*K. Wolk[1,¥], R.S. Dragland[1,¥], E. Chavez Panduro[1], L. Richarz[1], Z. Yan[2], E. Bourret[2], K.A. Hunnestad[1], Ch. Tzschaschel[3,4], J. Schultheiß[1,*], D. Meier[1,*]*

[1] Department of Materials Science and Engineering, Norwegian University of Science and Technology (NTNU), 7034 Trondheim, Norway
[2] Materials Sciences Division, Lawrence Berkeley National Laboratory, California 94720, USA
[3] Max Born Institute for Nonlinear Optics and Short Pulse Spectroscopy, Berlin 12489, Germany
[4] Department of Materials, ETH Zurich, 8092 Zurich, Switzerland
* jan.schultheiss@ntnu.no, dennis.meier@ntnu.no
¥ these authors contributed equally to the work.

**Engineering of ferroelectric domain structures enables direct control over the switching dynamics and is crucial for tuning the functional properties of ferroelectrics for various applications, ranging from capacitors to future nanoelectronics. Here, we investigate domain formation in poly- and single crystalline improper ferroelectric $DyMnO_3$. We show that a non-uniform grain-size distribution in the polycrystals facilitates the coexistence of multi-scale domains, varying by up to one order of magnitude in size. This unusual domain structure originates from an inverted domain-size/grain-size dependence that is intrinsic to the hexagonal manganite polycrystals, expanding previous studies towards non-uniform grain-size distributions. Our results demonstrate that the micrometer-sized grains in $DyMnO_3$ represent individual ferroelectric units with a characteristic domain structure, giving a new dimension to domain engineering in ferroelectric polycrystals with non-uniform microstructures.**

Ferroelectric materials exhibit a wide range of functional properties, such as piezoelectricity and high permittivity, as well as electrocaloric and electrooptic effects, playing an essential role for technological applications.[1] Below the Curie temperature, ferroelectrics develop a spontaneous electric polarization.[2] Depending on the underlying driving mechanism, they can be categorized into two main classes, i.e., proper and improper ferroelectrics. In proper ferroelectrics, the spontaneous polarization is the primary symmetry-breaking order parameter, arising from the displacement of ions away from their centrosymmetric positions, with well-known examples including $BaTiO_3$, $Pb(Zr,Ti)O_3$, and $(K,Na)NbO_3$. In improper ferroelectrics, the polarization emerges as a secondary effect, driven by, e.g., long-range magnetic order or a structural instability.[3,4] This improper nature also manifests on the level of domains, providing new opportunities for property engineering at nano- to macroscopic length scales.[5]
A well-established model system for improper ferroelectricity are the hexagonal manganites, $R$MnO$_3$ ($R$ = Sc, Y, In, and Dy-Lu). Here, the primary symmetry-breaking order parameter is a unit-cell tripling distortive mode, which leads to the formation of six structural trimerization domains.[6,7] In these domains, the polarization is either parallel (+$P$) or antiparallel (-$P$) to the hexagonal $c$-axis. The unusual distribution of ferroelectric 180° domains and their exotic properties are discussed elsewhere.[8-10] Recent investigations of $ErMnO_3$ polycrystals revealed conceptually new routes for the engineering of improper ferroelectric domains via mechanical pressure[11] and novel domain scaling effects in response to elastic confinement, leading to an inverted domain-size/grain-size dependency.[12]

In this work, we expand upon the research on domain physics in improper ferroelectric single- and polycrystalline materials towards $DyMnO_3$. In comparison to polycrystalline $ErMnO_3$, which is stable in the hexagonal $P6_3cm$ phase, $DyMnO_3$ lies at the stability edge between the hexagonal $P6_3cm$ and the orthorhombic $Pbnm$ phase[13] with largely unexplored consequences for the microstructure and emergent ferroelectric domains. We start with a comparison of key structural and ferroelectric properties of $DyMnO_3$ single- and polycrystals. We find that polycrystalline hexagonal $DyMnO_3$ features a highly non-uniform microstructure with grain sizes ranging from a few micrometers to several tens of micrometers. This variability is attributed to the system's proximity to the orthorhombic phase and associated differences in unit cell volume. Importantly, the size of the ferroelectric domains follows the non-uniform grain-size distribution, showing an inverted scaling behavior between domain and grain size. Larger grains feature smaller domains, which is consistent with previous studies on a series of polycrystalline $ErMnO_3$ with a homogeneous grain size distribution.[12] This correlation leads to a multi-scale domain distribution controlled by the $DyMnO_3$ microstructure, giving new opportunities for domain engineering in ferroelectrics.

We begin our discussion with a $DyMnO_3$ single crystal and its ferroelectric domain structure. The $DyMnO_3$ single crystal is grown by the pressurized floating-zone technique following the same procedure as outlined for $ErMnO_3$ in ref. 15. The as-grown crystal ingots are displayed in the inset of Figure 1a. Representative Laue back-reflection data are shown in Figure 1a, confirming the single-crystalline nature, hexagonal crystal structure, and phase purity of our sample. The polished $DyMnO_3$ sample with a surface roughness of 1.52 nm (RMS value) is characterized by piezoresponse force microscopy (PFM) to image the ferroelectric domain structure as presented in Figure 1b. The PFM data is obtained on a sample with a thickness of ~0.5 mm, recorded with a peak-to-peak voltage of 10 V applied to the back-electrode at a frequency of 40.13 kHz. Dark and bright areas correspond to ferroelectric domains of opposite polarization orientation, *P*, as indicated by the arrows. The domains form an isotropic pattern with six-fold meeting points, showing the characteristic domain structure reported for other members of the *R*$MnO_3$ family, also consistent with previous studies on $DyMnO_3$ thin films.[16,17] This leads us to the conclusion that despite the vicinity to the stability edge of the hexagonal phase, $DyMnO_3$ exhibits qualitatively the same ferroelectric domains as other hexagonal *R*$MnO_3$ systems with smaller ionic sizes of the *R* atom.[18-20]

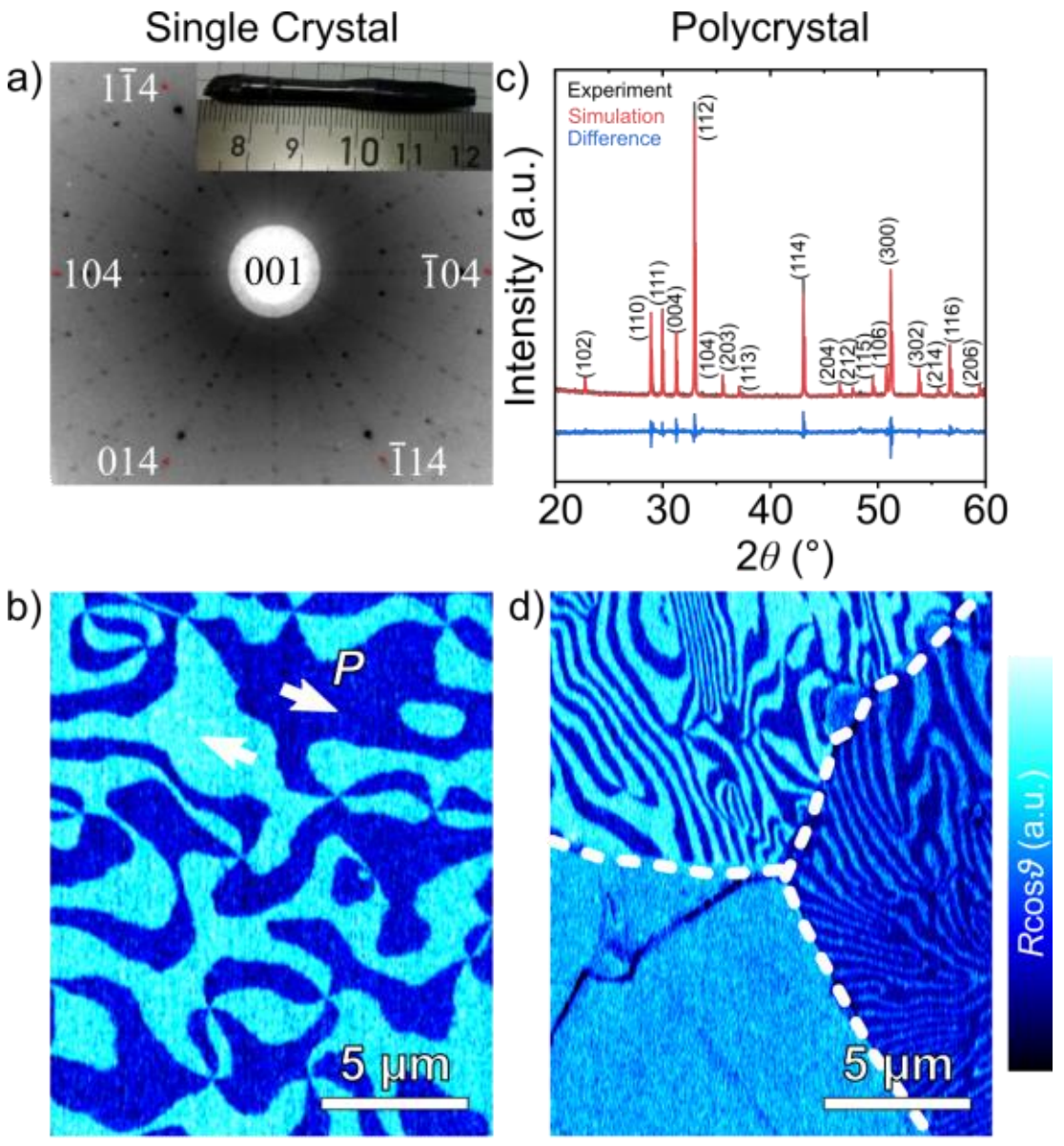

**Fig. 1:** Characterization of $DyMnO_3$ single and polycrystalline samples. a) Laue back-reflection data, indexed with the Cologne Laue Indexation Program[14], of a single crystalline $DyMnO_3$ sample. A photograph of a crystal ingot with a size of about 5 mm in diameter and 50 mm in length is displayed in the inset. b) In-plane piezoresponse contrast of $R \cdot \cos\vartheta$ ($R$ and $\vartheta$ refer to the amplitude and phase of the deflection of the laser signal, see method section), with the direction of the ferroelectric polarization, $P$, indicated by the arrows. The ferroelectric polarization is clearly distinguishable by the bright and dark contrast levels. c) XRD diffraction data of a polycrystalline ceramic pellet, heat-treated in argon atmosphere (1400°C, 12 hrs) with a hexagonal crystal structure. A Pawley fit (red curve) is fitted to the experimentally measured data (black curve). d) In-plane PFM contrast of polycrystalline $DyMnO_3$ of an area featuring three grains. The boundary between the grains is displayed by a dashed white line.

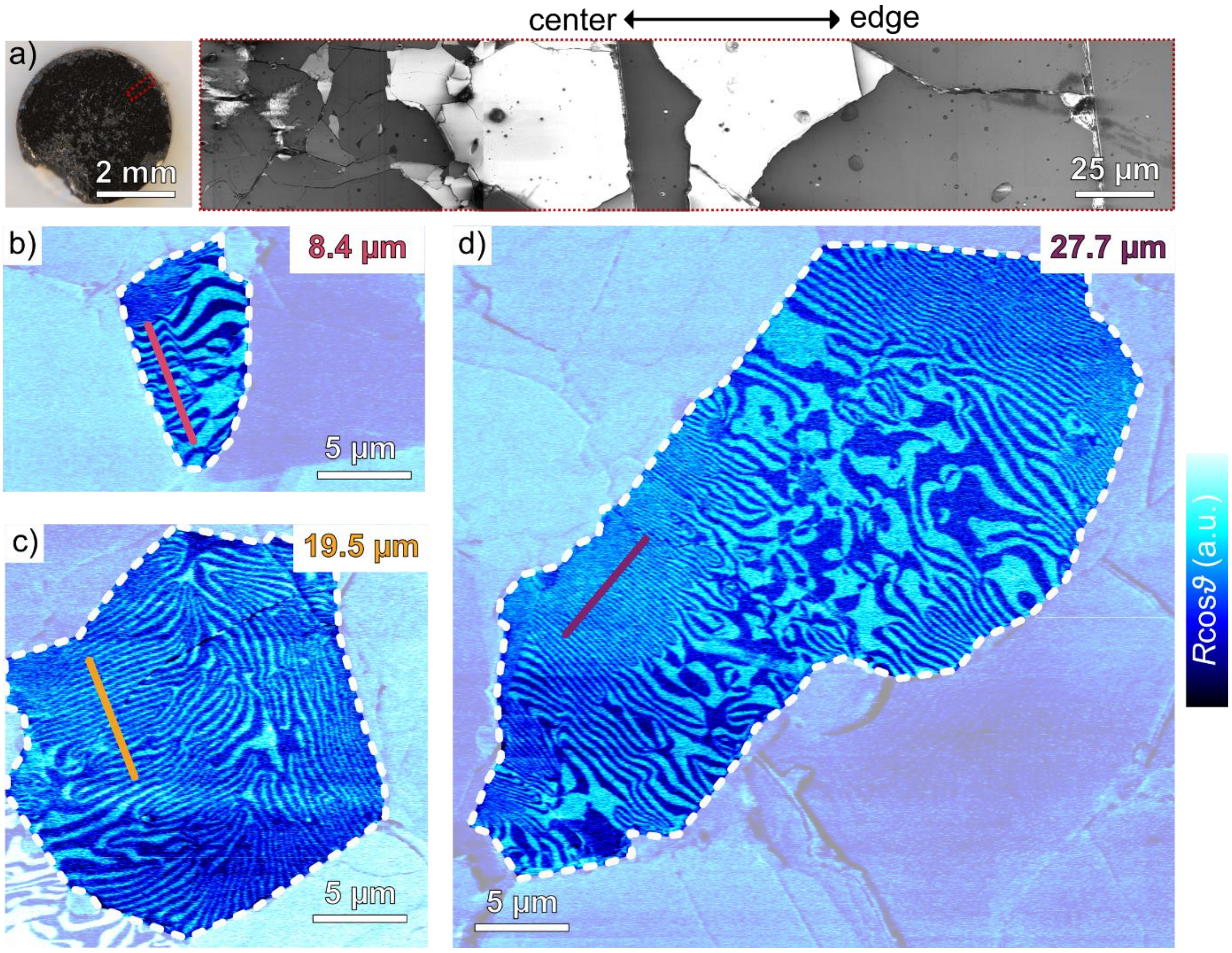

**Fig. 2:** a) SEM micrograph featuring a grain size gradient from large grain sizes at the edge towards smaller grain sizes in the center of the sample. An optical image of the sample is displayed on the left. b)-d) PFM data displaying the domain structures observed in grains of different sizes in polycrystalline $DyMnO_3$. a) 8.4 µm, b) 19.5 µm, and c) 27.7 µm. The white dotted lines visualize the position of the grain boundaries. The periodicity of the stripe-like domains was extracted along the solid lines and is displayed in Figure 3a.

Next, we move to $DyMnO_3$ polycrystals, studying the ferroelectric domain distribution in micro-structured systems. $DyMnO_3$ in the polycrystalline form is synthesized utilizing a solid-state reaction method (for processing details, see methods). In comparison to the synthesis of polycrystalline $ErMnO_3$, which can readily be grown in air atmosphere,[12] flowing argon is required to stabilize the hexagonal crystal structure in $DyMnO_3$ polycrystals during the final heat treatment step.[21] The phase-pure hexagonal target phase ($P6_3cm$) is verified by XRD as displayed in Figure 1c. The presence of several peaks at different angular values in the diffractogram confirms the polycrystalline nature of our sample. The lattice parameters of $a$ = 6.1843 Å and $c$ = 11.4592 Å, extracted by a Pawley fit[22] of the XRD data, are in agreement with previous structural investigations on hexagonal $DyMnO_3$ single crystals.[23,24] Polishing of the polycrystalline material results in a flat surface with a roughness of 4.97 nm (RMS value) within the grains. Corresponding PFM data are displayed in Figure 1d, showing domain contrast that varies in strength between grains as expected from the polycrystalline structure and literature data recorded on $ErMnO_3$ polycrystals.[11,12] Analogous to the $DyMnO_3$ single crystal (Figure 1b), two types of ferroelectric 180° domains with +*P* and -*P* polarization directions are observed. The domain pattern within the individual grains, however, is different from the one observed in the single crystal. Instead of an isotropic domain pattern, stripe-like domains are predominantly formed, which can be related to the presence of nonzero strain fields in the polycrystals,[25-27] as explained in ref. 12. Similar to polycrystalline hexagonal $ErMnO_3$,[12,28] we further find that the domain walls do not continue over the grain boundaries, which suggests that each grain represents an independent micrometer-sized ferroelectric entity regarding its ferroelectric order. The latter, in principle, gives an additional degree of freedom for domain engineering, allowing to generate coexisting grains with individual domain structures and, hence, individual electronic responses.

Going beyond the level of individual grains, we now turn to microstructure effects at larger length scales. Figure 2a shows a representative scanning electron microscopy (SEM) micrograph of a region of about 300 x 50 $µm^2$ as illustrated in the optical microscopy image on the left. The SEM data shows that the grains in our $DyMnO_3$ polycrystal are highly non-uniform in size. Towards the center, the grains have a size in the order of a few micrometers, whereas grains of tens of micrometer are observed at the sample edge. This is qualitatively different from the $ErMnO_3$ polycrystals reported earlier.[12] Non-uniform distributions of grain sizes in polycrystalline ceramics are a well-established phenomenon,[29,30] arising, for example, due to chemical trace impurities[31] or localized stresses[32] that hinder or promote grain growth in certain regions of the ceramic pellet. A possible origin for the non-uniform distribution of grain sizes we find in $DyMnO_3$ are impurities in the raw materials affecting local grain growth and/or the instability of $DyMnO_3$ towards different structural phases (i.e., hexagonal vs. orthorhombic with ~10 % volume change per formula unit[33]). In general, optimization of the processing allows for suppressing non-uniformity of grain size distributions. In this work, however, we follow a different strategy and utilize the non-uniform grain size distribution to learn more about the aforementioned decoupling of grains with respect to their ferroelectric order and explore related opportunities for domain engineering. Figure 2b-d presents PFM data revealing the ferroelectric domain structure in $DyMnO_3$ for grain sizes from 8.4 to 27.7 µm, showing distinct PFM contrast between the +*P* and -*P* domains and a mixture of isotropic and stripe-like domains.

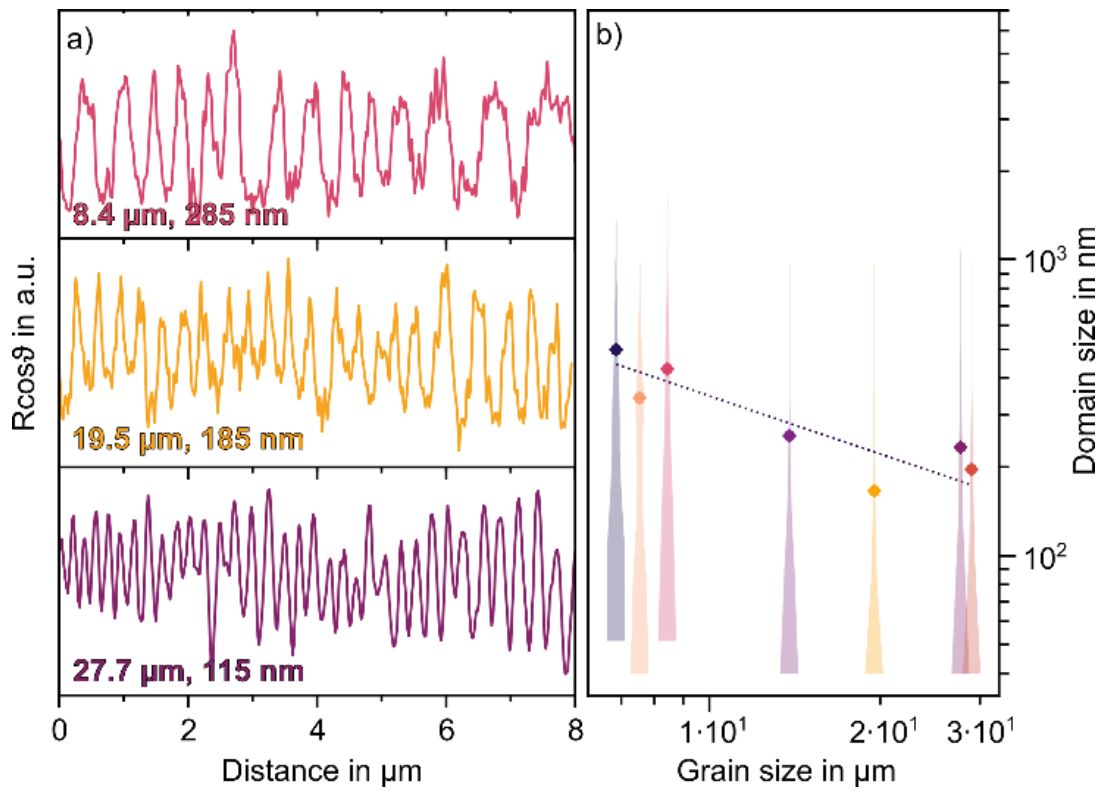

**Fig. 3:** a) Grain-size dependence of the domain size in polycrystalline $DyMnO_3$. Extracted cross sections of the piezoresponse along the colored lines in Figure 2 a)-c), showing a decrease in the periodicity of stipe-like domains with increasing grain size. b) Domain size as a function of the grain size in polycrystalline $DyMnO_3$. Data is obtained from grains of different sizes within the same polycrystal of $DyMnO_3$. A description of the analysis of domain sizes is provided in Figure S2. The distribution of the domain sizes is given together with the mean value. The dashed line represents a fit of the universal scaling law ($d \sim g^m$)[34] to the mean domain size with negative scaling exponent *m*.

To test for grain-size driven effects, we start by quantifying the periodicity of the stripe-like domain structure in selected regions for the three grains in Figure 2b-d. As displayed in Figure 3a, for polycrystalline $DyMnO_3$ samples with grain sizes from 8.4 to 27.7 µm, the periodicity of the stripe-like domains of the representative cross sections in Figure 2b-d decreases from 285 to 115 nm. It is important to note, however, that the difference in periodicity observed along the three test lines is not representative for the whole grain as it exhibits much more complex domain patterns with local variations in domain type and size. To account for this complexity and enhance the statistical robustness of our analysis, we employ a stereographical evaluation method to quantify the domain size as explained in detail in Figure S2. This approach leads to the violin plot in Figure 3b, summarizing the distribution of domain sizes for grains of different size. Overall, when considering the mean values in Figure 3b, a consistent trend is seen, indicating that the domain size, *d*, decreases with increasing grain size, *g*. The relation between *d* and *g* is commonly expressed by a power scaling law ($d \sim g^m$)[34], with *m* being negative in the case of $DyMnO_3$ as reflected by the black dotted line in Figure 3b. The negative scaling exponent indicates an inverted domain scaling behavior compared to proper ferroelectric materials,[35,36] consistent with previous experiments conducted on a series of polycrystalline $ErMnO_3$ samples with different grain sizes[12].

The observation of a negative *m* (domain scaling exponent) within one and the same polycrystalline sample with a non-uniform grain size distribution is remarkable as it shows the universality and robustness of the inverted scaling behavior in hexagonal manganites, applying despite substantial non-uniformity of the microstructure of the system. For each grain, the inverted grain-size/domain-size relationship holds, independent of the size of the neighboring grains. The latter reflects that the grains can indeed be considered as individual ferroelectric units, with independent ferroelectric domain structures. Thus, opposite to previous studies on proper ferroelectrics, such as $BaTiO_3$[37] or $(K,Na)NbO_3$[38], where non-uniform grain sizes are usually unwanted due to their negative impact on dielectric reliability[39] and densification behavior[40], our results present inhomogeneous grain size distribution as an additional degree of freedom for domain engineering.[41-43]

As demonstrated based on the model system $DyMnO_3$, coexisting multi-scale domains can be achieved within one and the same sample. The possibility to engineer such distributions is interesting, as they influence the dynamics of polarization reversal, giving new opportunities for tuning the temporal distribution of the electromechanical strain[44] and the magnitude of piezoelectric responses in ferroelectrics[45]. In general, non-uniform grain-size distributions naturally occur in a wide range of (improper) ferroelectric polycrystals – or can readily be induced on demand – providing an exciting pathway towards functional ferroic materials with enhanced complexity.

## Methods

**Synthesis.** The $DyMnO_3$ single crystal growth follows the same procedure described for $ErMnO_3$ single crystals in ref. 15. The $DyMnO_3$ ingots were oriented by Laue diffraction and cut into specimens with surfaces normal to the (110)-direction in order to facilitate microscopy experiments on the crystal surfaces with the ferroelectric polarization being in plane, *P*||(001). For the synthesis of the polycrystals, $Dy_2O_3$ (99.9% purity; Sigma-Aldrich, St. Louis, MO, USA) and $Mn_2O_3$ (99.0% purity; Sigma-Aldrich, St. Louis, MO, USA) were furnace dried at a temperature of 900°C and 700°C, respectively for a dwell time of 12 hrs. Stoichiometric amounts of both precursors were weighed, ball milled for 24 hrs at 205 rpm and stepwise heat treated. A first heat-treatment step was carried out at 900 °C in air for 12 hours. The powder XRD pattern is displayed in Supplementary Figure S1a, featuring unreacted educts as well as other reaction byproducts, such as $DyMn_2O_5$. A second heat-treatment step was carried out at 1100 °C in air for 12 hours, resulting in phase-pure $DyMnO_3$ with an orthorhombic crystal structure (*Pbnm*) with the XRD pattern displayed in Figure S1b. Subsequently, the powder was pressed uniaxially into cylindrical pellets with a diameter of 5 mm, followed by isostatic compaction under a pressure of 200 MPa. The final densification of the pressed pellet was performed in a flowing argon gas for 12 hrs (ETF 17 horizontal tube furnace, Entech, Ängelholm, Sweden). In agreement to previous reports[21] and confirmed by our XRD analysis (Figure 1c), the heat treatment step in argon gas resulted in the final hexagonal crystal structure ($P6_3cm$). A cooling and heating of 5°C/min was utilized for the sintered pellet.

**Crystallographic analysis.** The orientation of the single crystalline $DyMnO_3$ sample was determined by Laue XRD in back scattering geometry using a molybdenum X-ray tube. Diffraction data were recorded using photographic plates that were digitalized using a Fujifilm BAS image plate scanner and indexed with the Cologne Laue Indexation Program.[14] After orientation, crystals were cut into slabs with surfaces normal to the (110)-direction using a diamond wire saw. XRD patterns of mortared powders were obtained in between each firing step (XRD, D8 ADVANCE Davinci design, Bruker with Cu Kα radiation) to identify the crystallographic phases. A Pawley fit[22] using Topas[46] was performed to determine the lattice parameter.

**Microscopy.** Lapping of the single and the polycrystals was carried out utilizing a 9 µm-grained $Al_2O_3$ water suspension (Logitech Ltd, Glasgow, UK) followed by a subsequent polishing step utilizing a silica slurry (SF1 Polishing Fluid, Logitech AS, Glasgow, Scotland). PFM was performed using an NT-MDT system (NT-MDT, Moscow, Russia) with an electrically conductive tip (Spark 150 Pt, Nu Nano Ltd., Bristol, UK). The deflection of the laser signal was read out as the amplitude, *R*, and the phase, $\vartheta$, of the piezoresponse using two lock-in amplifiers (SR830, Stanford Research Systems, CA, USA). The assignment of the polarization direction and contrast level was achieved by calibration of the PFM system using a periodically poled $LiNbO_3$ reference sample (PFM03, NT-MDT, Moscow, Russia). A Helios NanoLab DualBeam Focused Ion Beam system (Thermo Fisher Scientific, MA, USA) was used for the SEM imaging, using secondary electrons as the imaging source detected with an in-lens detector. The acceleration voltage was set to 5 kV, and the beam current to 0.4 nA.

## Acknowledgements

The authors thank J. Koruza for helpful discussions. K. W. acknowledges financial support provided by the Erasmus+ program for the traineeship grant. J.S. acknowledges support from the Alexander von Humboldt Foundation through the Feodor-Lynen fellowship. D.M. thanks NTNU for support through the Onsager Fellowship Program and the outstanding Academic Fellow Program. R.S.D., L.R., and D.M. acknowledge funding from the European Research Council (ERC) under the European Union's Horizon 2020 Research and Innovation Program (Grant Agreement No. 863691). The authors acknowledge financial support from the NTNU Nano Impact fund. The Research Council of Norway (RCN) is acknowledged for the support to the Norwegian Micro- and Nano-Fabrication Facility, NorFab (project number 295864).

## Data availability

The data that supports the findings of this study are available from the corresponding author upon reasonable request.

# Supporting Information

## Coexistence of multi-scale domains in ferroelectric polycrystals with non-uniform grain-size distributions

K. Wolk[1,¥], R.S. Dragland[1,¥], E. Chavez Panduro[1], L. Richarz[1], Z. Yan[2], E. Bourret[2], K.A. Hunnestad[1], Ch. Tzschaschel [3,4], J. Schultheiß[1,*], D. Meier[1,*]

[1] Department of Materials Science and Engineering, Norwegian University of Science and Technology (NTNU), 7034 Trondheim, Norway

[2] Materials Sciences Division, Lawrence Berkeley National Laboratory, California 94720, USA

[3] Max Born Institute for Nonlinear Optics and Short Pulse Spectroscopy, Berlin 12489, Germany

[4] Department of Materials, ETH Zurich, 8092 Zurich, Switzerland

* jan.schultheiss@ntnu.no, dennis.meier@ntnu.no

[¥]these authors contributed equally to the work.

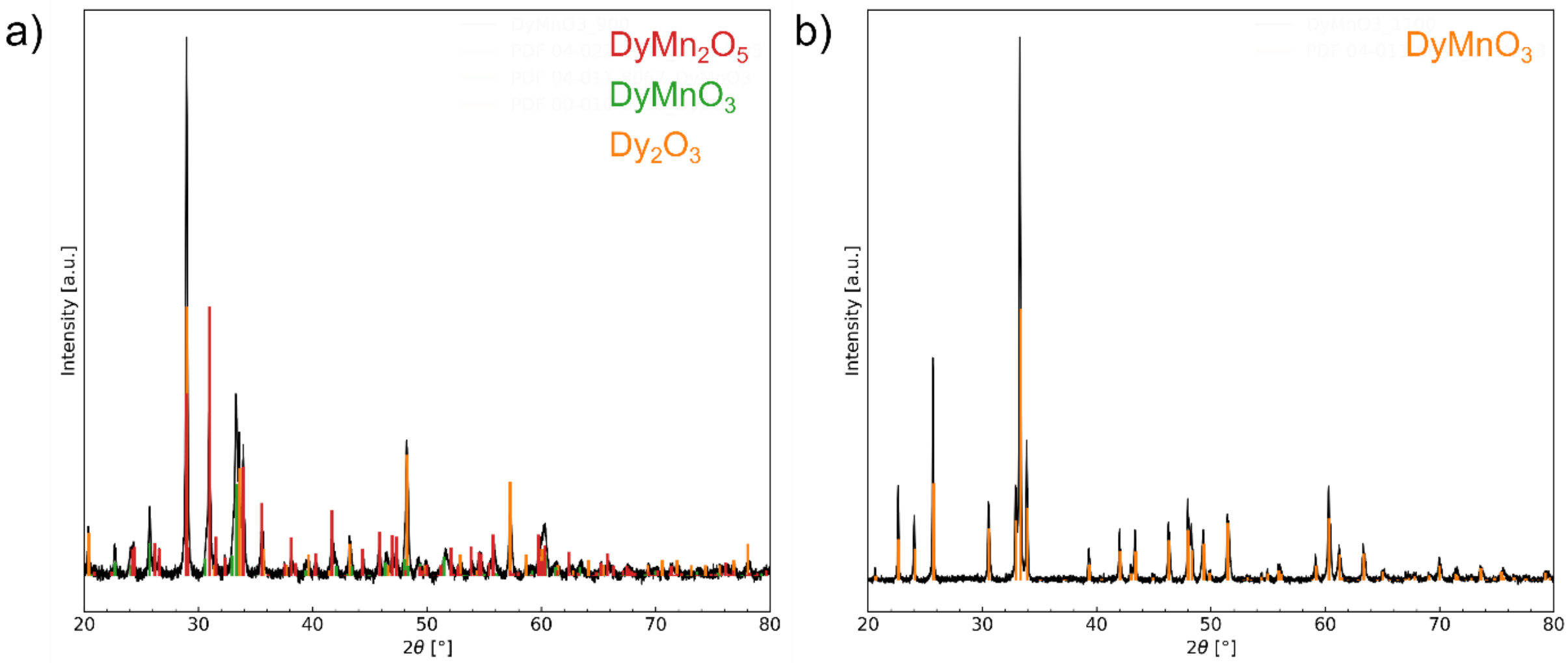

**Figure S1:** XRD pattern of stoichiometric mixtures of $Mn_2O_3$ and $Dy_2O_3$ raw materials under different heat treatment conditions. The powders are first heat treated at a) 900°C, 12 hrs, air and subsequently at b) 1100°C, 12 hrs, air. The diffraction peaks are indicated as $DyMn_2O_5$ (PDF 04-022-2995), orthorhombic $DyMnO_3$ (PDF 04-011-8067), and $Dy_2O_5$ (PDF 00-010-0059). As it is visible from the XRD patterns, the subsequent heat-treatment step results in a phase pure $DyMnO_3$ phase.

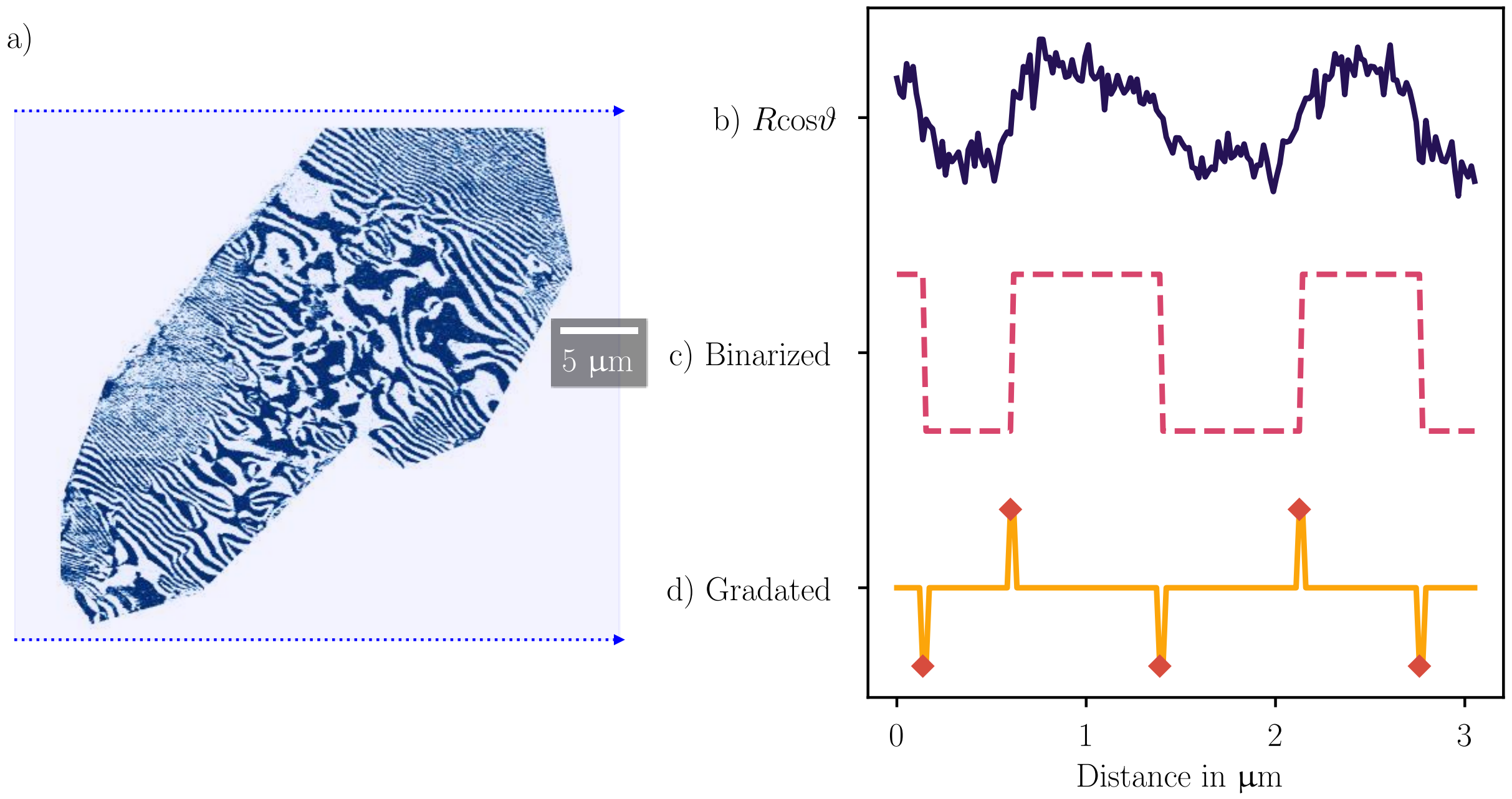

**Figure S2:** Illustration of the grain and domain size estimation algorithms used to obtain the quantitative relationship displayed in Figure 3b. The grain boundaries are identified manually, as illustrated by the PFM contrast for a representative grain in a). First, a Boolean value of 1 is assigned to every pixel within these boundaries. Next, the grain size is estimated as the diameter of a circle within the derived area. For domain size estimation, the mentioned Boolean mask is initially used to remove the signal surrounding the grain. Subsequently, the image is rotated by a given angle, and every pixel outside the grain is set to NaN (undefined). An independent NaN-excluding average is found along every pixel row, with the raw PFM data, $R\cos\vartheta$, displayed for a representative cross section in b). Each pixel row is then binarized against the line's respective average value. The resulting binarized profile for one such line is displayed in c). Finally, the gradients of the binarized profiles displayed in d) reveal the domain boundaries, indicated by red diamonds. All distances between two subsequent domain boundaries are then retrieved for each pixel row, contributing to the domain size distribution of the grain. Upon completion, the procedure is repeated for every 3° rotation angle, to achieve sufficient angular sampling.